\documentclass[a4paper,11pt]{article}
\usepackage{lineno}
\usepackage{amssymb}
\usepackage{amsfonts,amsbsy,graphics,epsfig}
\usepackage{graphicx}
\usepackage{amsmath}
\usepackage{color}
\usepackage{multirow}
\usepackage{arydshln}

\usepackage[dvipsnames]{xcolor}
\usepackage{wasysym}
\usepackage{natbib}
\usepackage{cleveref}
\usepackage{adjustbox}
\modulolinenumbers[1]

\usepackage{stmaryrd} 
\usepackage{dsfont} 

\begin{document}
\begin{center}
%
%

\textbf{\Large{Investigating spatial scan statistics for multivariate functional data}}\medskip \\
Camille Frévent$^1$, Mohamed-Salem Ahmed$^{1}$ , Sophie Dabo-Niang$^2$ and Michaël Genin$^1$  \\
$^1$Univ. Lille, CHU Lille, ULR 2694 - METRICS: Évaluation des technologies de santé et des pratiques médicales, F-59000 Lille, France.\\ 
$^2$Laboratoire Paul Painvelé UMR CNRS 8524, INRIA‐MODAL, University of Lille.
\end{center}	
\begin{center}
	\rule{1\linewidth}{.9pt}
\end{center}
\noindent	
\textbf{\Large Abstract}\medskip \\
\noindent This paper introduces new scan statistics for multivariate functional data indexed in space. The new methods are derivated from a MANOVA test statistic for functional data, an adaptation of the Hotelling $T^2$-test statistic, and a multivariate extension of the Wilcoxon rank-sum test statistic. In a simulation study, the latter two methods present very good performances and the adaptation of the functional MANOVA also shows good performances for a normal distribution. Our methods detect more accurate spatial clusters than an existing nonparametric functional scan statistic. Lastly we applied the methods on multivariate functional data to search for spatial clusters of abnormal daily concentrations of air pollutants in the north of France in May and June 2020.\\

\begin{center}
\rule{1\linewidth}{.9pt}
\end{center}
\noindent \section{Introduction}

Spatial cluster detection has been studied for many years. The goal is usually to develop new tools capable of detecting the aggregation of spatial sites that behave ``differently'' from other sites. In particular, spatial scan statistics detect statistically significant spatial clusters with a scanning window and without any pre-selection bias. This approach was originally proposed by \cite{spatialdisease} and \cite{spatialscanstat} in the cases of Bernouilli and Poisson models. They present a method based on the likelihood ratio and Monte-Carlo testing to detect significant clusters of various sizes and shapes. Following on from Kulldorff's initial work, several researchers have adapted spatial scan statistics to other spatial data distributions, such as ordinal \citep{jung2007spatial}, normal \citep{normalkulldorff}, exponential \citep{huang2007spatial}, and Weibull models \citep{bhatt2014spatial}. These methods were applied in many different fields such as epidemiology \citep{epidemio1, epidemio2, genin2020fine}, environmental sciences \citep{social2, social1}, geology \citep{geology}. \\

\noindent Thanks to progress in sensing and data storage capacity, data are increasingly being measured continuously
over time.
This led to the introduction of functional data analysis (FDA) by \cite{ramsaylivre}. A considerable amount of work has gone to adapt classical statistical methods to the univariate functional framework such as principal component analysis \citep{fpc_boente, fpc_berrendero} or regression \citep{reg_cuevas, reg_ferraty, reg_chiou} but also to the multivariate functional one \citep{local_functional, clustering_multi}). \\

\noindent In some research fields, such as environmental surveillance, pollution sensors are deployed in a geographical area. In a context where these sensors measure simultaneously the concentrations of many pollutants at regular intervals over a long period of time, environmental experts may search for environmental black-spots
, that can be defined as geographical areas characterized by elevated concentrations of pollutants. For this purpose three different approaches can be considered. The simplest one consists in summarizing the information by averaging each variable over the time and to apply a parametric multivariate spatial scan statistic \citep{a_multivariate_gaussian} or the nonparametric one proposed by \cite{nonparam_multi} but this could lead to a huge loss of information when the data is measured over a long time period. Another solution could be to apply  a spatial scan statistic for univariate functional data on each variable \citep{wilco_cucala, notre_fonctionnel}. However this does not allow to take into account the correlations between the variables. A relevant solution consists in using spatial scan statistics for multivariate functional data. According to the authors, the nonparametric spatial scan statistic for functional data proposed by \cite{wilco_cucala} could be extended to multivariate processes although it has never been evaluated in this context. Moreover to our knowledge no parametric scan statistic for multivariate functional data has been proposed.
Thus we will define new spatial scan statistics for multivariate functional data based on statistical tests for comparing multivariate functional samples. Recently \cite{manovafonctional} and \cite{two_s} developed respectively a MANOVA test statistic and a functional Hotelling $T^2$-test statistic for multivariate functional data. We also propose to consider the multivariate extension of the Wilcoxon rank-sum test developed by \cite{oja} as a pointwise test statistic. Using these statistics we will adapt to the multivariate functional framework the parametric and the distribution-free spatial scan statistics for functional data proposed by \cite{notre_fonctionnel} and we will also investigate a new multivariate functional method based on the ranks of the observations at each time.
\\

\noindent This paper develops three new spatial scan statistics for multivariate functional data.
Section \ref{sec:method} describes the parametric multivariate functional scan statistic, the multivariate version of the distribution-free functional spatial scan statistic proposed by \cite{notre_fonctionnel} and a new rank-based spatial scan statistic for multivariate functional data
. In Section \ref{sec:simulation} the behaviours of our methods are investigated through a simulation study and compared to the one proposed by \cite{wilco_cucala}. The methods are applied on a real dataset in Section \ref{sec:realdata}. Finally the paper is concluded with a discussion in Section \ref{sec:discussion}.

\section{Methodology} \label{sec:method}

\subsection{General principle}
\noindent Let $\{ X(t), \ t \in \mathcal{T} \}$ be a $p$-dimensional vector-valued stochastic process where $\mathcal{T}$ is an interval of $\mathbb{R}$. Let $s_1, \dots, s_n$ be $n$ non-overlapping locations of an observation domain $S \subset \mathbb{R}^2$ and $X_1, \dots, X_n$ be the observations of $X$ in $s_1, \dots, s_n$. Hereafter all observations are considered to be independent, which is a classical assumption in scan statistics. \noindent Spatial scan statistics aim at detecting spatial clusters and testing their significance. Hence, one tests a null hypothesis $\mathcal{H}_0$ (the absence of a cluster) against a composite alternative hypothesis $\mathcal{H}_1$ (the presence of at least one cluster $w \subset S$ presenting abnormal values of $X$).

\noindent \cite{notre_fonctionnel} defined  the notion of cluster in the univariate functional framework. Their 
definitions can be easily extended to the multivariate functional context by defining a 
\textit{multivariate magnitude cluster} $w$ as follows:
\begin{equation}
\forall t \in \mathcal{T}, \ \mathbb{E}[X_i(t)\mid s_i\in w] = \mathbb{E}[X_i(t)\mid s_i\notin w] + \Delta(t),
\end{equation}
where $\Delta(t) = (\Delta_1(t), \dots, \Delta_p(t))^\top$, all $\Delta_i$ are of constant and identical signs, and exists $i \in \llbracket 1 ; p \rrbracket$ such that $\Delta_i$ is non-zero over at least one sub-interval of $\mathcal{T}$.
In the same way a \textit{multivariate shape cluster} can be defined as follows:
\begin{equation}
\forall t \in \mathcal{T}, \ \mathbb{E}[X_i(t)\mid s_i\in w] = \mathbb{E}[X_i(t)\mid s_i\notin w] + \Delta(t)
\end{equation}
where $\Delta(t) = (\Delta_1(t), \dots, \Delta_p(t))^\top$ and exists $i \in \llbracket 1 ; p \rrbracket$ such that $\Delta_i$ is not constant almost everywhere. \\

\noindent Since the article of \cite{cressie}, a scan statistic is defined by the maximum of a concentration index over a set of potential clusters $\mathcal{W}$. 
In the following and without loss a generality, we focus on  variable-size circular clusters  \citep[as introduced by][]{spatialscanstat}. The set of potential clusters $\mathcal{W}$ is the set of discs centered on a location and passing through another one, with $|w|$ the number of sites in $w$:
\begin{equation}
\mathcal{W} = \{ w_{i,j} \ / \ 1 \le |w_{i,j}| \le \frac{n}{2}, \ 1 \le i,j \le n \},
\end{equation}
where $w_{i,j}$ is the disc centered on $s_i$ that passes through $s_j$. Thus, a cluster cannot cover more than 50\% of the studied region which is the recommended approach of \cite{spatialdisease}. 
Remark that in the literature other possibilities have been proposed such as elliptical clusters \citep{elliptic}, rectangular clusters \citep{rectangular} or graph-based clusters \citep{cucala_graph}. \\

\noindent We proposed a parametric scan statistic in subsection \ref{subsec:param}, a distribution-free one is detailed in subsection \ref{subsec:dfree} and a new rank-based scan statistic for multivariate functional data is developed in subsection \ref{subsec:max_wmw}.

\subsection{A parametric spatial scan statistic for multivariate functional data} \label{subsec:param}

\noindent In this subsection the process $X$ is supposed to take values in the Hilbert space $L^2(\mathcal{T}, \mathbb{R}^p)$ of $p$-dimensional vector-valued square-integrable functions on $\mathcal{T}$, equipped with the inner product $\langle X, Y \rangle = \int_{\mathcal{T}} X(t)^\top Y(t) \ \text{d}t$.

\noindent \cite{notre_fonctionnel} proposed a parametric scan statistic for univariate functional data based on a functional ANOVA. A multivariate version of the ANOVA is the classical MANOVA Lawley–Hotelling trace test \citep{oja}. It was adapted by \cite{manovafonctional} for $L^2(\mathcal{T}, \mathbb{R}^p)$ processes:
considering two groups $g_1$ and $g_2$ of independent random observations of two $p$-dimensional stochastic processes $X_{g_1}$ and $X_{g_2}$ taking values in 
$L^2(\mathcal{T}, \mathbb{R}^p)$, it tests the equality of the two mean vector-valued functions $\mu_{g_1}$ and $\mu_{g_2}$
where $\mu_{g_i}(t) = \mathbb{E}[X_{g_i}(t)] \in \mathbb{R}^p$, $i=1,2, \ t \in \mathcal{T}$.  \\

\noindent For the cluster detection problem, the null hypothesis $\mathcal{H}_0$ (the absence of a cluster) can be defined by $\mathcal{H}_0: \forall w \in \mathcal{W}, \ \mu_{w} = \mu_{w^\mathsf{c}} = \mu_S$, where $\mu_w$, $\mu_{w^\mathsf{c}}$ and $\mu_S$ stand for the mean functions in $w$, outside $w$ and over $S$, respectively. And the alternative hypothesis $\mathcal{H}_1^{(w)}$ associated with a potential cluster $w$ can be defined as $\mathcal{H}_1^{(w)}: \mu_{w} \neq \mu_{w^\mathsf{c}}$. Thus we can use the functional MANOVA to compare the mean functions in $w$ and $w^\mathsf{c}$. \\
Actually \cite{manovafonctional} presented the adaptation of different MANOVA tests to the functional framework. However the Wilks lambda test statistic, the Lawley-Hotelling trace test statistic and the Pillai trace test statistic presented in the article showed similar performances. In addition, they often outperformed in terms of power the tests proposed in the same article that use projections. Thus we decide to study the Lawley-Hotelling trace test for the cluster detection problem by using the following statistic:
\begin{equation}
\mathrm{LH}^{(w)} = \text{Trace}(H_w E_w^{-1}) \end{equation}
where $$H_w = |w| \int_{\mathcal{T}} [\bar{X}_w(t) - \bar{X}(t)][\bar{X}_w(t) - \bar{X}(t)]^\top \ \text{d}t + |w^\mathsf{c}| \int_{\mathcal{T}} [\bar{X}_{w^\mathsf{c}}(t) - \bar{X}(t)][\bar{X}_{w^\mathsf{c}}(t) - \bar{X}(t)]^\top \ \text{d}t$$ 
\text{ and }
$$E_w = \sum_{j, s_j \in w} \int_{\mathcal{T}} [X_j(t) - \bar{X}_w(t)][X_j(t) - \bar{X}_w(t)]^\top \text{d}t + \sum_{j, s_j \in w^\mathsf{c}} \int_{\mathcal{T}} [X_j(t) - \bar{X}_{w^\mathsf{c}}(t)][X_j(t) - \bar{X}_{w^\mathsf{c}}(t)]^\top \text{d}t$$
where $\bar{X}_{g}(t) = \frac{1}{|g|} \sum_{i, s_i \in g} X_i(t)$ are empirical estimators of $\mu_g(t)$ ($g \in \{w, w^\mathsf{c}\}$), $\bar{X}(t) = \frac{1}{n} \sum_{i = 1}^n X_i(t)$ is the empirical estimator of $\mu_S(t)$. \\

\noindent Now, $\mathrm{LH}^{(w)}$ can be considered as a concentration index and maximized over the set of potential clusters $\mathcal{W}$, which results in the following definition of the parametric multivariate functional spatial scan statistic (PMFSS):
\begin{equation}
\Lambda_{\text{PMFSS}} = \underset{w \in \mathcal{W}}{\max} \  \mathrm{LH}^{(w)}.
\end{equation}
The potential cluster for which this maximum is obtained, namely the most likely cluster (MLC) is 
\begin{equation}
\text{MLC} = \underset{w \in \mathcal{W}}{\arg \max} \  \mathrm{LH}^{(w)}.
\end{equation}

\subsection{A distribution-free spatial scan statistic for multivariate functional data} \label{subsec:dfree}

\cite{notre_fonctionnel} proposed a distribution-free spatial scan statistic for univariate functional data based on the combination of the distribution-free scan statistic for non-functional data proposed by \cite{a_distribution_free} which relies on a Student's t-test, and the globalization of a pointwise test over the time \citep{hd_manova}. \\
Very recently, \cite{two_s} proposed a version of this pointwise test for $p$-dimensional functional data ($p \ge 2$) to compare the mean functions of $X$ in two groups. \\

\noindent We suppose that for each time $t$, $\mathbb{V}[X_i(t)] = \Sigma(t,t)$ for all $i \in \llbracket 1 ; n \rrbracket$, where $\Sigma$ is a $p \times p$ covariance matrix function. \\

\noindent Thus, as previously, in the context of cluster detection, the null hypothesis $\mathcal{H}_0$ can be defined as follows: $\mathcal{H}_0: \forall w \in \mathcal{W}, \ \mu_{w} = \mu_{w^\mathsf{c}} = \mu_S$, where $\mu_w$, $\mu_{w^\mathsf{c}}$ and $\mu_S$ stand for the mean functions in $w$, outside $w$ and over $S$, respectively. And the alternative hypothesis $\mathcal{H}_1^{(w)}$ associated with a potential cluster $w$ can be defined as follows: $\mathcal{H}_1^{(w)}: \mu_{w} \neq \mu_{w^\mathsf{c}}$. Next, \cite{two_s} proposed to compare the mean function $\mu_w$ in $w$ with the mean function  $\mu_{w^\mathsf{c}}$ in $w^\mathsf{c}$ by using the following statistic:
$$ T_{n,\text{max}}^{(w)} = \underset{t \in \mathcal{T}}{\sup} \ T_n(t)^{(w)}$$
where $T_n(t)$ is a pointwise statistic defined by the Hotelling $T^2$-test statistic
$$ T_n(t)^{(w)} = \frac{|w| |w^\mathsf{c}|}{n} (\bar{X}_w(t) - \bar{X}_{w^\mathsf{c}}(t))^\top \hat{\Sigma}(t,t)^{-1} (\bar{X}_w(t) - \bar{X}_{w^\mathsf{c}}(t)).$$
$\bar{X}_w(t)$ and $\bar{X}_{w^\mathsf{c}}(t)$ are the empirical estimators of the mean functions defined in subsection \ref{subsec:param}, and $$\hat{\Sigma}(s,t) = \frac{1}{n-2} \left[ \sum_{i, s_i \in w} (X_i(s) - \bar{X}_w(s)) (X_i(t) - \bar{X}_w(t))^\top + \sum_{i, s_i \in w^\mathsf{c}} (X_i(s) - \bar{X}_{w^\mathsf{c}}(s)) (X_i(t) - \bar{X}_{w^\mathsf{c}}(t))^\top \right]$$ is the pooled sample covariance matrix
function. \\

\noindent Then $T_{n,\text{max}}^{(w)}$ is considered as a concentration index and maximized over the set of potential clusters $\mathcal{W}$, yielding to the following multivariate distribution-free functional spatial scan statistic (MDFFSS):

\begin{equation}
\Lambda_{\text{MDFFSS}} = \underset{w \in \mathcal{W}}{\max} \  T_{n,\text{max}}^{(w)}.
\end{equation}
The most likely cluster is therefore
\begin{equation}
\text{MLC} = \underset{w \in \mathcal{W}}{\arg \max} \  T_{n,\text{max}}^{(w)}.
\end{equation}

\subsection{A new rank-based spatial scan statistic for multivariate functional data} \label{subsec:max_wmw}

\cite{oja} developed a $p$-dimensional ($p \ge 2$) extension of the classical Wilcoxon rank-sum test using multivariate ranks. Following on \cite{oja}'s definitions, we can define the notion of ``pointwise multivariate ranks'' as following: \\
For each time $t \in \mathcal{T}$, the pointwise
multivariate ranks are defined by
$$R_i(t) = \frac{1}{n} \sum_{j=1}^{n} \text{sgn}(A_X(t)(X_i(t) - X_j(t)))$$ where 
$\text{sgn}(\cdot)$ is the spatial sign function defined as 
$$\begin{array}{ccccl}
\text{sgn} &: & \mathbb{R}^p & \to & \mathbb{R}^p \\
 & & x & \mapsto & \left\{
\begin{array}{cl}
  ||x||_2^{-1} x & \text{ if } x \neq 0 \\
  0 & \text{ otherwise}
\end{array}
\right. \\
\end{array}$$
and 
$A_X(t)$ is a pointwise data-based transformation matrix that makes the pointwise multivariate ranks behave as
though they are spherically distributed in the unit $p$-sphere:
$$ 
\frac{p}{n}\sum_{i=1}^{n} R_i(t) R_i(t)^\top = \frac{1}{n} \sum_{i=1}^{n} R_i(t)^\top R_i(t) I_p.$$ 
Note that this matrix can be easily computed using an iterative procedure. \\

\noindent Without loss of generality, \cite{oja} 
compared the cumulative distribution functions of real multivariate observations in two groups. In the context of multivariate functional data, their statistic can be considered as a pointwise test statistic for each time $t$: the pointwise multivariate extension of the Wilcoxon rank-sum test statistic is defined as

$$W(t)^{(w)} = \frac{pn}{\sum_{i=1}^{n} R_i(t)^\top R_i(t)} \left[ \ |w| \ ||\bar{R}_w(t) ||_2^2 + |w^\mathsf{c}| \  ||\bar{R}_{w^\mathsf{c}}(t) ||_2^2 \ \right]$$ where $\bar{R}_{g}(t) = \frac{1}{|g|} \sum_{i, s_i \in g} R_i(t) $ for $g \in \{ w, w^\mathsf{c} \}.$ \\

\noindent Now we propose as previously to globalize the information over the time with $$W^{(w)} = \underset{t \in \mathcal{T}}{\sup} \ W(t)^{(w)}.$$

\noindent Then in the context of cluster detection, the null hypothesis is defined as $\mathcal{H}_0$: $\forall w \in \mathcal{W}, \ \forall t, \ F_{w,t} = F_{w^\mathsf{c},t}$ where $F_{w,t}$ and $F_{w^\mathsf{c},t}$ correspond respectively to the cumulative distribution functions of $X(t)$ in $w$ and outside $w$. The alternative hypothesis $\mathcal{H}_1^{(w)}$ associated with a potential cluster $w$ is $\mathcal{H}_1^{(w)}$: $\exists t$, \  $F_{w,t}(x) = F_{w^\mathsf{c},t}(x-\Delta_t)$, $\Delta_t \neq 0$. \\

\noindent Then $W^{(w)}$ can be considered as a concentration index and maximized over the set of potential clusters $\mathcal{W}$ so that the multivariate rank-based functional spatial scan statistic (MRBFSS) is defined as follows:

\begin{equation}
\Lambda_{\text{MRBFSS}} = \underset{w \in \mathcal{W}}{\max} \  W^{(w)}.
\end{equation}
Thus the most likely cluster is
\begin{equation}
\text{MLC} = \underset{w \in \mathcal{W}}{\arg \max} \  W^{(w)}.
\end{equation}

\subsection{Computing the significance of the MLC} \label{subsec:significance}

\noindent Once the most likely cluster has been detected, its significance must be evaluated. 
The distribution of the scan statistic $\Lambda$ ($\Lambda_{\text{PMFSS}}$, $\Lambda_{\text{MDFFSS}}$ or $\Lambda_{\text{MRBFSS}}$) is untractable under $\mathcal{H}_0$ due to the dependence between $\mathcal{S}^{(w)}$ and $\mathcal{S}^{(w')}$ if $w \cap w' \neq \emptyset$ ($\mathcal{S} = \mathrm{LH}, T_{n,\text{max}}$ or $W$). 
Then we chose to obtain a large set of simulated datasets by randomly permuting the observations $X_i$ in the spatial locations. This technique called ``random labelling'' was already used in spatial scan statistics \citep{normalkulldorff, a_multivariate_gaussian, notre_fonctionnel}.

\noindent Let $M$ denote the number of random permutations of the original dataset and $\Lambda^{(1)},\dots,\Lambda^{(M)}$ be the observed scan statistics on the simulated datasets. According to \cite{dwass} the p-value for $\Lambda$ observed in the real data is estimated by 
\begin{equation}
\hat{p} = \frac{1 + \sum_{m=1}^M \mathds{1}_{\Lambda^{(m)} \ge \Lambda}}{M+1}.
\end{equation}
Finally, the MLC is considered to be statistically significant if the associated $\hat{p}$ is less than the type I error. \\

\section{A simulation study} \label{sec:simulation}
\noindent 
\noindent A simulation study was conducted to compare the performances of the parametric multivariate functional spatial scan statistic (PMFSS) $\Lambda_{\text{PMFSS}}$, the multivariate distribution-free functional spatial scan statistic (MDFFSS) $\Lambda_{\text{MDFFSS}}$ and the new multivariate rank-based functional spatial scan statistic (MRBFSS) $\Lambda_{\text{MRBFSS}}$. \cite{wilco_cucala} proposed a nonparametric scan statistic for univariate functional data (NPFSS) $\Lambda_{\text{NPFSS}}$. However according to the authors it can be extended to the multivariate functional framework although it has not been studied in this context. Thus we decided to include their approach in the simulation, using the computation improvement proposed by \cite{notre_fonctionnel}.

\subsection{Design of simulation study}

\noindent Artificial datasets were generated by using the geographic locations of the 94 French \textit{départements} (county-type administrative areas) as shown in Figure \ref{fig:sitesgrid}. The location of each \textit{département} was defined by its administrative capital. For each artificial dataset, a spatial cluster $w$ (composed of eight \textit{départements} in the Paris region ; the red area, see Figure \ref{fig:sitesgrid} in Supplementary materials) was simulated.

\subsubsection{Generation of the artificial datasets}

 \noindent The $X_i$ were simulated according to the following model with $p = 2$  \citep[see][for more details]{two_s, article_milano}:
\begin{center} 
$\text{for each }i \in \llbracket 1 ; 94 \rrbracket,\  X_i(t) = (\sin{[2\pi t^2]}^5 ; 1 + 2.3 t + 3.4 t^2 + 1.5 t^3)^\top + \Delta(t) \mathds{1}_{s_i \in w} + \varepsilon_{i}(t), \ t \in [0;1]. $
\end{center}
where $\varepsilon_{i}(t) = \sum_{k=1}^{100} Z_{i,k} \sqrt{1.5 \times 0.2^k} \theta_k(t)$, with $\theta_k(t) = \left\{ \begin{array}{ll}
1 & \text{ if } k = 1 \\
\sqrt{2} \sin{[k\pi t]} & \text{ if } k \text{ even} \\
\sqrt{2} \cos{[(k-1)\pi t]} & \text{ if } k \text{ odd and } k > 1
\end{array} \right.$. \\

\noindent The functions $X_i$ were measured at 101 equally spaced times on $[0;1]$. \\

\noindent Giving $\Sigma = \begin{pmatrix} 1 & \rho \\ \rho & 1 \end{pmatrix}$ the covariance matrix of the $Z_{i,k}$, three distributions for the $Z_{i,k}$ were considered: (i) a normal distribution: $Z_{i,k} \sim \mathcal{N}(0,\Sigma)$, (ii) a standardized Student distribution: $Z_{i,k} = U_{i,k} \left( \frac{V_{i,4}}{4} \right)^{-0.5}$ where the  $U_{i,k}$ are independent $\mathcal{N}(0, \Sigma/2)$ variables and the $V_{i,4}$ are independent $\chi_2(4)$ variables and (iii) a standardized chi-square distribution: $Z_{i,k} = \left[ U_{i,k} - \begin{pmatrix} 4 \\ 4 \end{pmatrix} \right] / (2\sqrt{2})$ where the $U_{i,k}$ are independent and $U_{i,k} \sim \chi_2(4,\Sigma) = \Gamma(2,1/2, \Sigma)$ (rate parameterization).
Remark that $\rho$ is also the correlation of the two components of $X(t)$ for each time. \\

\noindent Three values of correlation $\rho$ were tested: $\rho = 0.2$, $0.5$ and $0.8$, and three types of clusters with intensity controlled by some parameter $\alpha > 0$ were studied: $\Delta_1(t) = \alpha (t ; t)^\top$, $\Delta_2(t) = \alpha ( t (1-t) ; t (1-t) )^\top$ and $\Delta_3(t) = \alpha (\exp{[ - 100 (t-0.5)^2 ]}/3 ; \exp{[ - 100 (t-0.5)^2 ]}/3)^\top$. Since they vary over time and are positive and non-zero on $\mathcal{T} = [0;1]$ (except possibly in $t = 0$ or $t = 1$), they correspond to both multivariate magnitude and multivariate shape clusters. \\ 
Different values of the parameter $\alpha$ were considered for each $\Delta$: $\alpha \in \{0 ; \ 0.375 ; \ 0.75 ; \ 1.125 ; \ 1.5 \}$ for $\Delta_1$, $\alpha \in \{0 ; \ 1 ; \ 2 ; \ 3 ; \ 4 \}$ for $\Delta_2$ and $\alpha \in \{ 0 ; \ 1.25 ; \ 2.5 ; \ 3.75 ; \ 5 \}$ for $\Delta_3$. 
Note that $\alpha = 0$ was also tested in order to evaluate the maintenance of the nominal type I error. An example of the data for $\rho = 0.2$ and for the Gaussian distribution for the $Z_{i,k}$ is given in the Appendix (Figure \ref{fig:examplesimu}).

\subsubsection{Comparison of the methods}

\noindent For each distribution of the $Z_{i,k}$, each type of $\Delta$, each level of correlation $\rho$, and each value of $\alpha$, 1000 artificial datasets were simulated. The type I error was set to 5\% and 999 samples were generated by random permutations of the data to evaluate the p-value associated with each MLC.
The performances of the methods were compared through four criteria: the power, the true positive rate, the false positive rate and the F-measure. \\ 

\noindent The power was estimated by the proportion of simulations yielding to the rejection of $\mathcal{H}_0$ according to the type I error. Among the simulated datasets yielding to the rejection of $\mathcal{H}_0$, the true positive rate is the average proportion of sites correctly detected among the sites in $w$, the false positive rate is the average proportion of sites in $w^\mathsf{c}$ that were included in the detected cluster and the F-measure corresponds to the average harmonic mean of the proportion of sites in $w$ within the detected cluster (positive predictive value) and the true positive rate.

\subsubsection{Results of the simulation study}

\begin{figure}[h!]
\centering
\includegraphics[width=\linewidth]{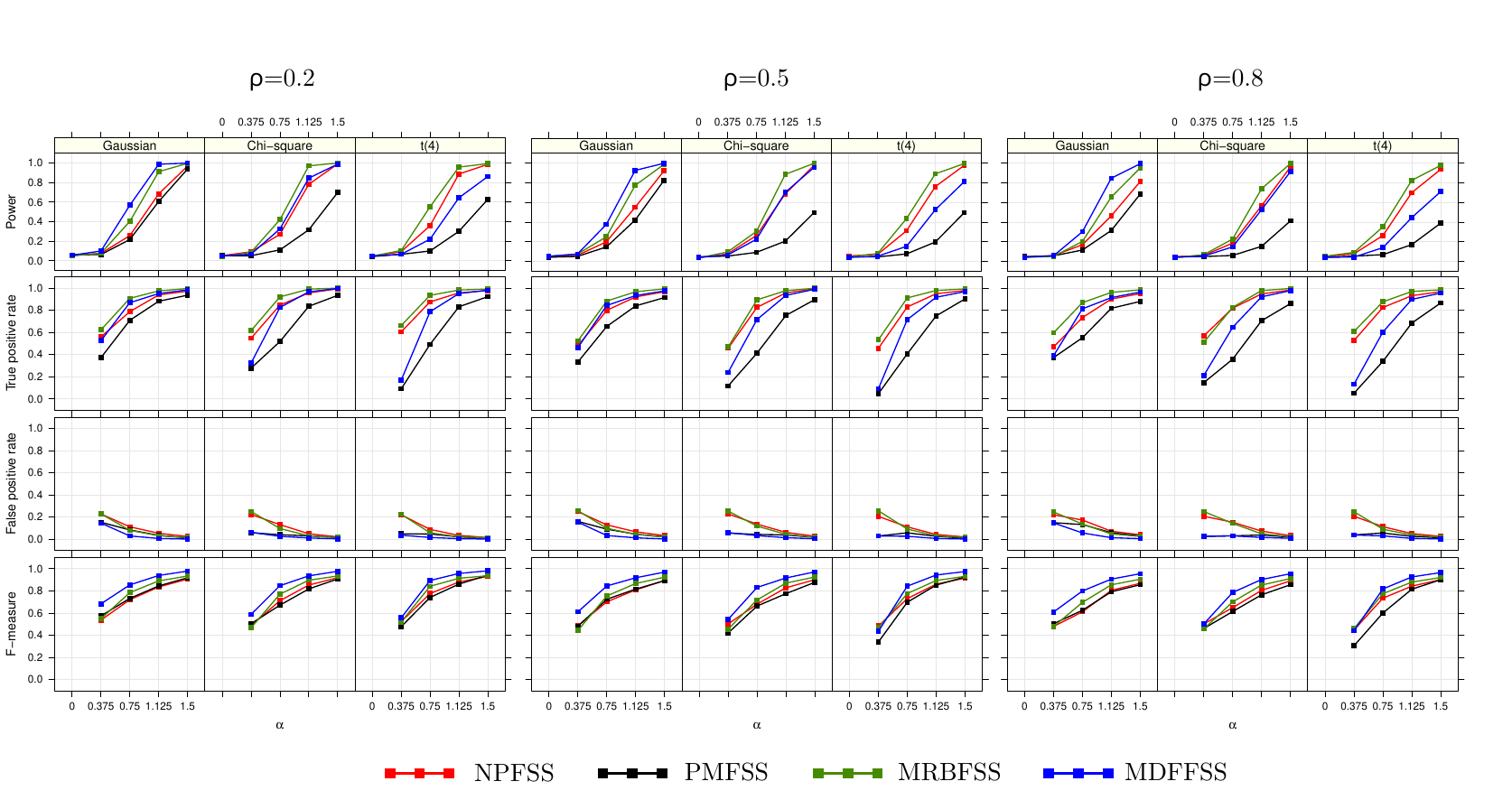}
\caption{The simulation study: comparison of the NPFSS, MDFFSS, MRBFSS and PMFSS methods for the shift $\Delta_1(t) = (\alpha t ; \alpha t)^\top$. For each method and each level of correlation $\rho$, the power curves, the true positive and false positive rates, and the F-measure values for detection of the spatial cluster as the MLC are shown. $\alpha$ is the parameter that controls the cluster intensity.}
\label{fig:resultssimu1}
\end{figure}

\begin{figure}[h!]
\centering
\includegraphics[width=\linewidth]{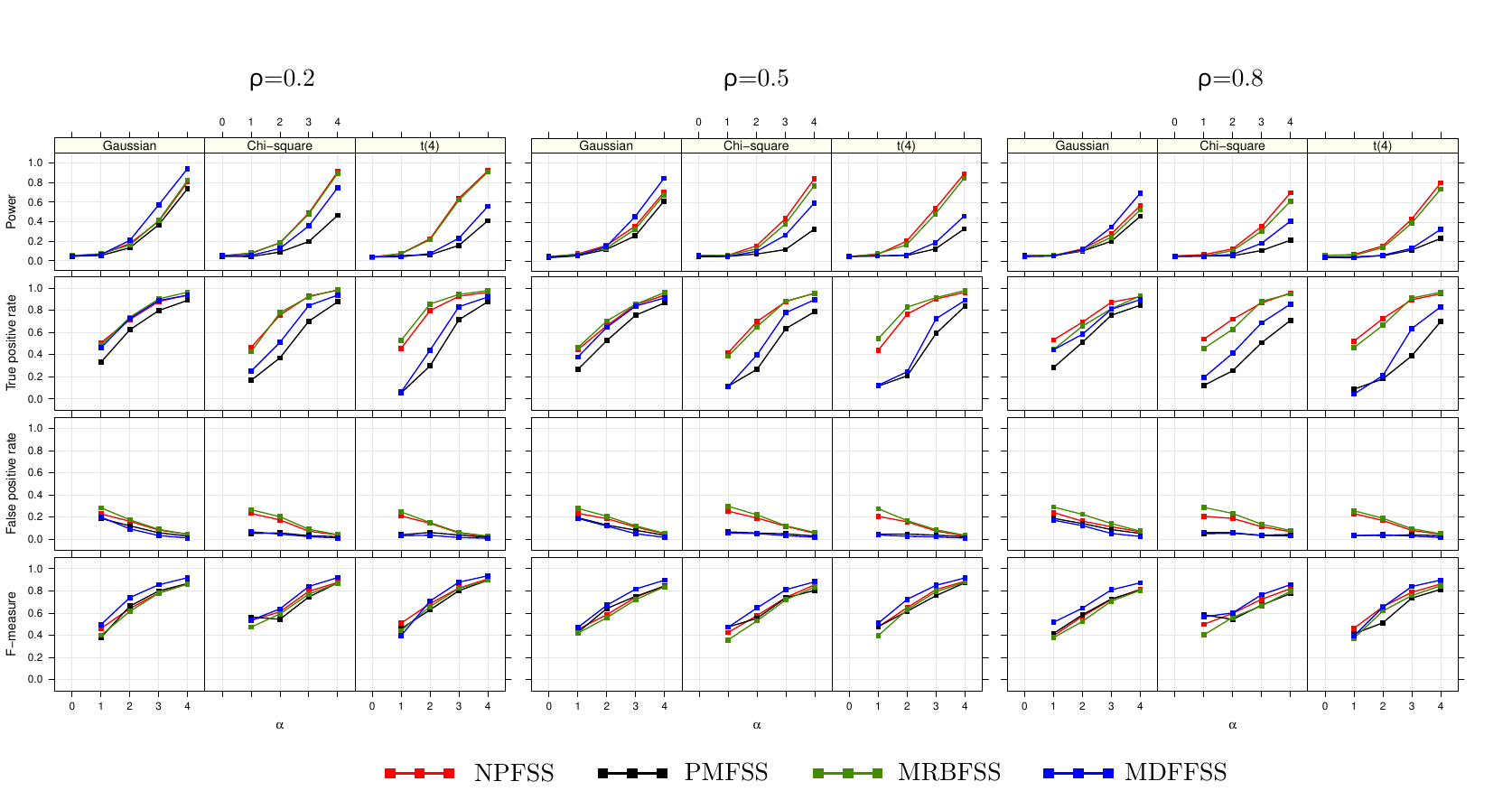}
\caption{The simulation study: comparison of the NPFSS, MDFFSS, MRBFSS and PMFSS methods for the shift $\Delta_2(t) = (\alpha t (1-t) ; \alpha t (1-t))^\top$. For each method and each level of correlation $\rho$, the power curves, the true positive and false positive rates, and the F-measure values for detection of the spatial cluster as the MLC are shown. $\alpha$ is the parameter that controls the cluster intensity.}
\label{fig:resultssimu2}
\end{figure}

\begin{figure}[h!]
\centering
\includegraphics[width=\linewidth]{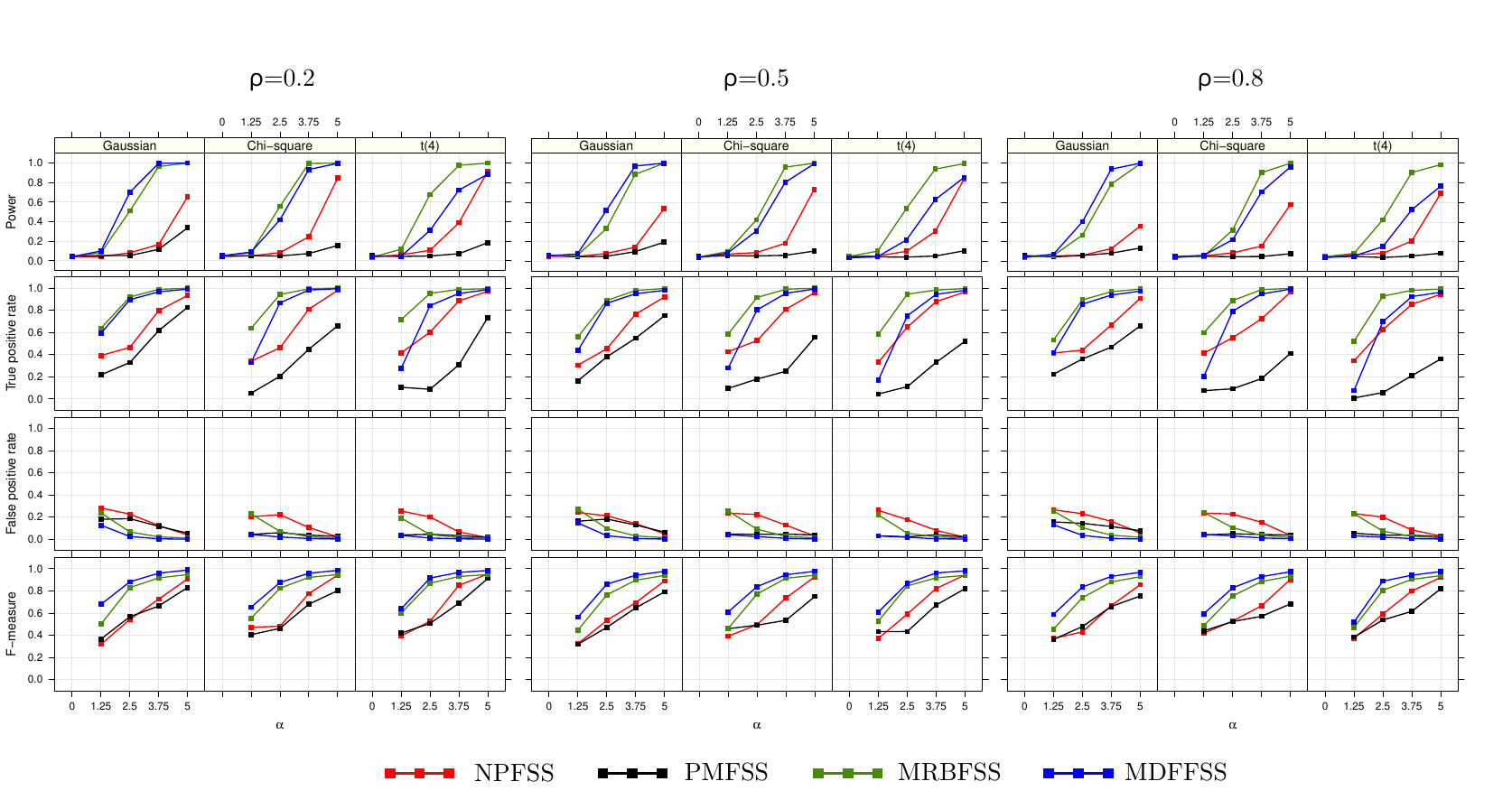}
\caption{The simulation study: comparison of the NPFSS, MDFFSS, MRBFSS and PMFSS methods for the shift $\Delta_3(t) = (\alpha \exp{[ - 100 (t-0.5)^2 ]}/3 ; \alpha \exp{[ - 100 (t-0.5)^2 ]}/3)^\top$. For each method and each level of correlation $\rho$, the power curves, the true positive and false positive rates, and the F-measure values for detection of the spatial cluster as the MLC are shown. $\alpha$ is the parameter that controls the cluster intensity.}
\label{fig:resultssimu3}
\end{figure}

\noindent The results of the simulation are presented in Figures \ref{fig:resultssimu1}, \ref{fig:resultssimu2} and \ref{fig:resultssimu3}. \\
\noindent For $\alpha = 0$, all methods seem to maintain the correct type I error of 0.05 regardless of the type of process, the type of $\Delta$ and the level of correlation $\rho$ (see the power curves in Figures \ref{fig:resultssimu1}, \ref{fig:resultssimu2} and \ref{fig:resultssimu3}). \\

\noindent For all methods the performances slightly decrease when the correlation $\rho$ increases. \\
\noindent The NPFSS and the PMFSS show similar powers for the Gaussian distribution for the shifts $\Delta_1$ and $\Delta_2$. However for non-Gaussian distributions of the $Z_{i,k}$ or the shift $\Delta_3$, the NPFSS presents higher powers than the PMFSS. The MDFFSS presents the highest powers in the Gaussian case. However its performances also decrease when the data are not distributed normally: in that case the MRBFSS shows the highest powers (except for $\Delta_2$ even if they are still very high). In the Gaussian case it also presents better powers than the NPFSS (except for $\Delta_2$) and the PMFSS. \\
\noindent The MRBFSS almost always shows the highest true positive rates (except sometimes for $\Delta_2$). The true positive rates for the MDFFSS are also very high for normal data but they decrease for non-normal data. The PMFSS presents the lowest true positive rates, however it presents very low false positive rates. 
In terms of false positives, the MDFFSS always shows the better performances. The MRBFSS often shows higher false positive rates, however they are lower than the ones of the NPFSS (except for $\Delta_2$ although both are very close). As a result the MDFFSS shows the highest F-measures, followed by the MRBFSS. For $\Delta_1$, in the Gaussian case the PMFSS and the NPFSS present similar F-measures whereas the F-measures are lower for the PMFSS for non-normal distributions. The NPFSS, the PMFSS and the MRBFSS shows very close F-measures for the shift $\Delta_2$ and finally the F-measures for the NPFSS and the PMFSS are strongly lower than the ones of the MDFFSS and the MRBFSS for the local shift $\Delta_3$.

\section{Application on real data} \label{sec:realdata}

\subsection{Air pollution in \textit{Nord-Pas-de-Calais}}

\begin{figure}[h!]
\centering
\includegraphics[width = \textwidth]{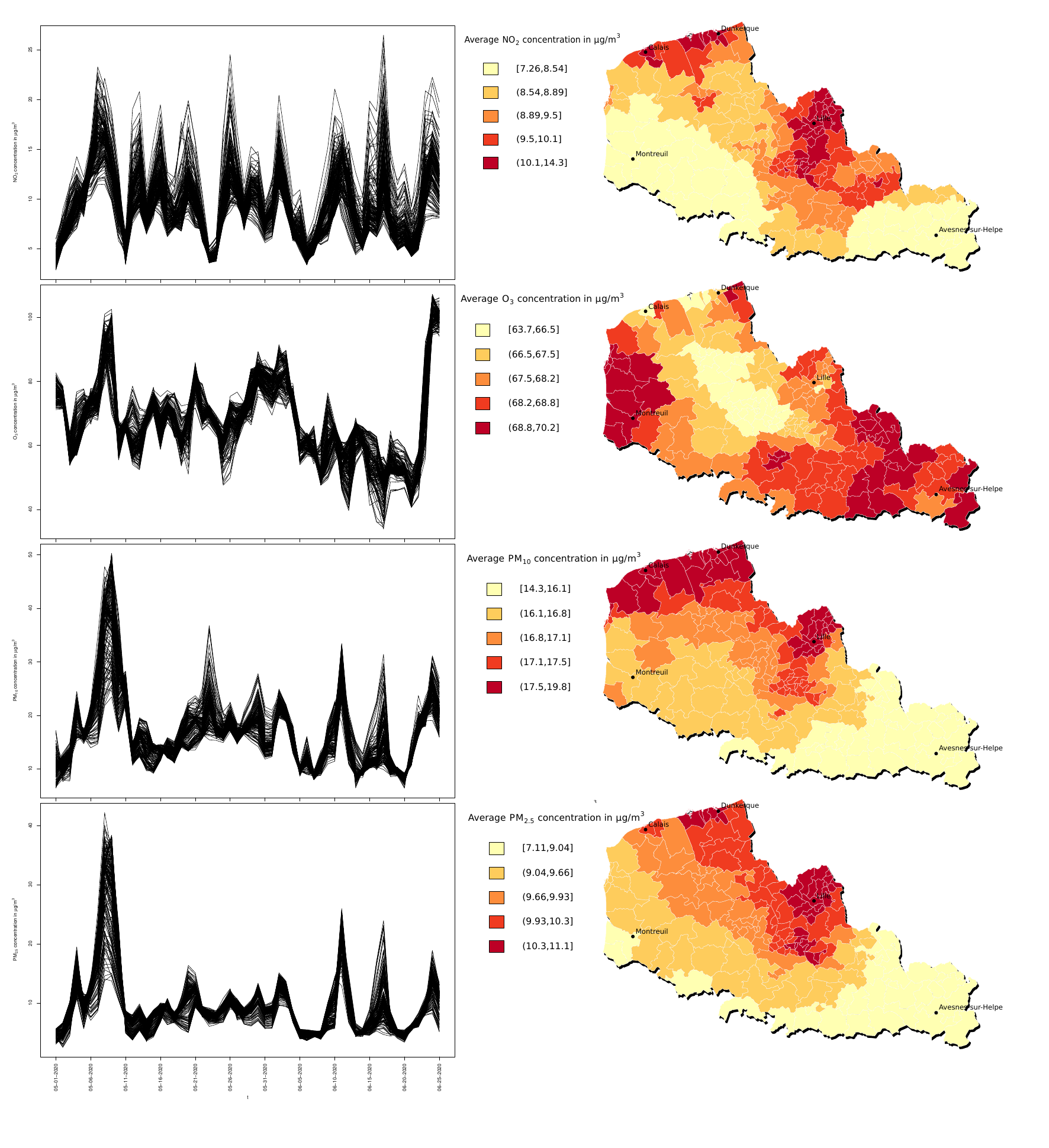}
\caption{Daily concentration curves of $\text{NO}_2$, $\text{O}_3$, $\text{PM}_{10}$ and $\text{PM}_{2.5}$ (from May 1, 2020 to June 25, 2020) in each of the 169 \textit{cantons} of \textit{Nord-Pas-de-Calais} (a region in northern France) (left panels), and the spatial distributions of the average concentrations for each pollutant over period from from May 1, 2020 to June 25, 2020 (right panels).}
\label{fig:courbes_pollutions}
\end{figure}

\noindent The data considered is the concentration in $\mu g.m^{-3}$ of four pollutants: ozone ($\text{O}_3$), nitrogen dioxide ($\text{NO}_2$) and fine particles $\text{PM}_{10}$ and $\text{PM}_{2.5}$ corresponding respectively to particles whose diameter is less than $10 \mu m$ and $2.5 \mu m$. Note that the $\text{PM}_{2.5}$ particles are included in the $\text{PM}_{10}$ particles. The data provided by the French national air quality forecasting platform PREV'AIR consists in the daily average of these variables from May 1, 2020 to June 25, 2020 (56 values for each variable) aggregated at the \textit{canton} (administrative subdivisions of \textit{départements}) level for each of the 169 \textit{cantons} of the \textit{Nord-Pas-de-Calais} (a region in northern France) located by their center of gravity.
The pollutants daily concentration curves in each \textit{canton} are presented in Figure \ref{fig:courbes_pollutions} (left panels) and the spatial distributions of the average concentrations for each pollutant over the studied time period are presented in Figure \ref{fig:courbes_pollutions} (right panels).

\noindent The maps in Figure \ref{fig:courbes_pollutions} show a spatial heterogeneity of the average concentration for each pollutant. High concentrations of $\text{O}_3$ tend to aggregate in the rural areas of Montreuil and Avesnes-sur-Helpe, and high concentrations of the other pollutants tend to aggregate in the urban areas of Calais, Dunkerque and Lille. Moreover the daily concentration curves present a marked temporal variability during the period from May 1, 2020 to June 25, 2020. Thus functional spatial scan statistics seem relevant to highlight the presence of \textit{cantons}-level spatial clusters of pollutants concentrations.

\subsection{Spatial clusters detection}
\noindent For the sake of concision, we have decided to present here only one method based on the results of the simulation. With regard to the latter, we have chosen the MRBFSS because it presents stable performances whatever the correlation and the distribution of the variables.

\noindent 
We considered a round-shaped scanning window of maximum radius 10 km since small clusters of pollution are more relevant for interpretation because the sources of the pollutants are very localized: the main source of $\text{NO}_2$ is road traffic, and $\text{PM}_{2.5}$ is mainly emitted in urban (heating, road traffic) or industrial areas. 
The statistical significance of the MLC was evaluated through 999 Monte-Carlo permutations and the MLC is said to be statistically significant if its p-value is less than 0.05.
Cluster detection was also performed for the other three methods, the results are presented in Figure \ref{fig:othermethods} and Table \ref{table:noconstraint} in Supplementary materials.

\subsection{Results}

\begin{figure}[h!]
\centering
\includegraphics[width = 1\textwidth]{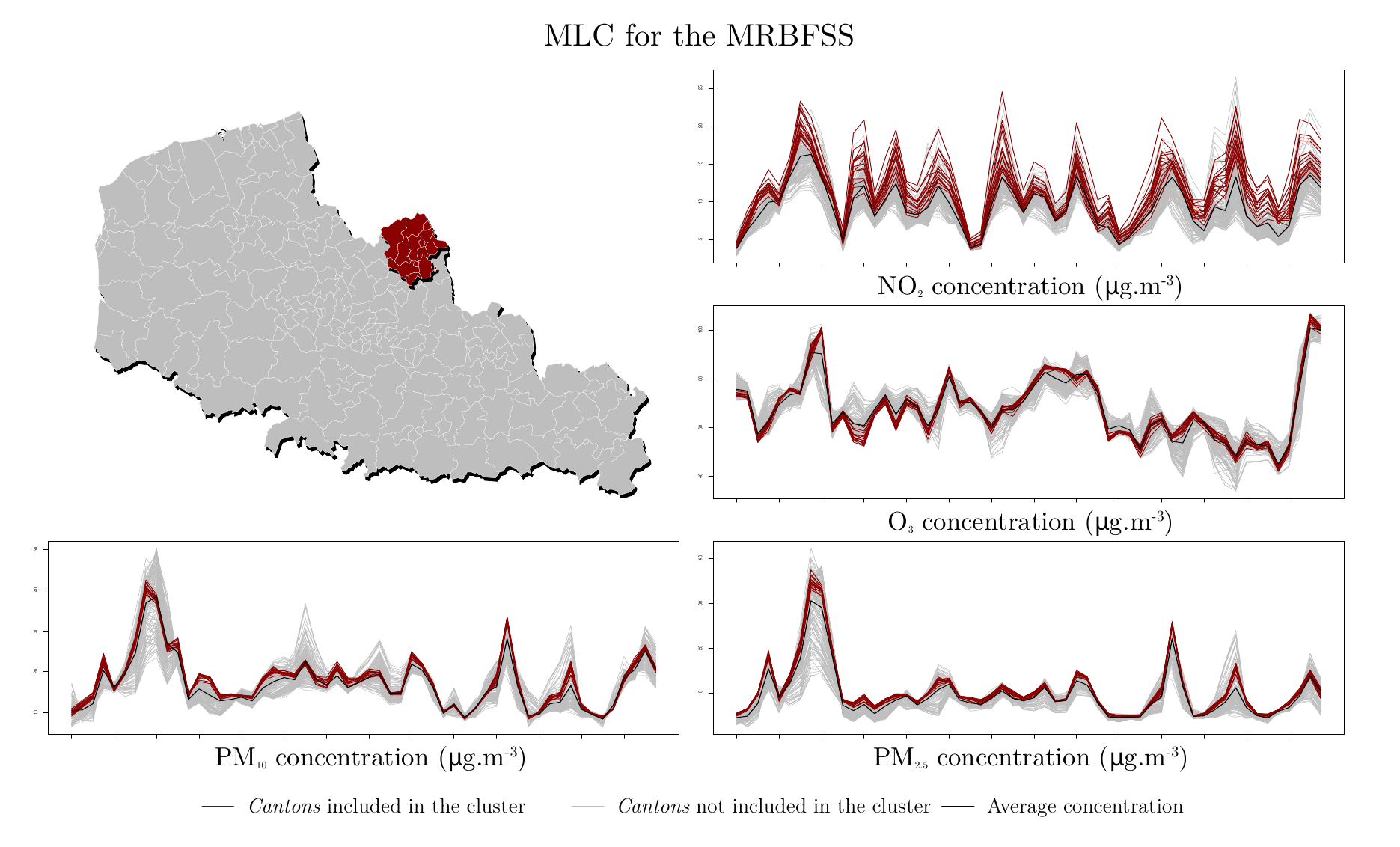}

\caption{Most likely cluster of pollutants ($\text{NO}_2$, $\text{O}_3$, $\text{PM}_{10}$ and $\text{PM}_{2.5}$) concentrations detected by the MRBFSS. The daily concentration curves of the pollutants (from May 1, 2020 to June 25, 2020) in each \textit{canton} are presented with colored lines. The black curves are the daily average concentration curves in the \textit{Nord-Pas-de-Calais} (a region in northern France).}
\label{fig:noconstraint}
\end{figure}

\noindent The MRBFSS detected a significant most likely cluster (15 \textit{cantons}, $308\text{km}^2$, $\hat{p} = 0.001$) in the area of Lille. This cluster, corresponding to high values of $\text{NO}_2$, $\text{PM}_{10}$ and $\text{PM}_{2.5}$ concentrations, is presented in Figure \ref{fig:noconstraint}. For these three pollutants all the curves in the MLC are above the average concentrations in the \textit{Nord-Pas-de-Calais}. 
In environmental science it is well-known that those pollutants are more frequent in urban areas. Therefore this is consistent with the cluster observed here.

\section{Discussion} \label{sec:discussion}

\noindent Here we developed a parametric multivariate functional scan statistic (PMFSS), a multivariate distribution-free functional spatial scan statistic (MDFFSS) and a multivariate rank-based functional spatial scan statistic (MRBFSS) which allow to detect clusters of abnormal values on multivariate functional data indexed in space. The goal of such methods is to alert the scientists if abnormal values are detected. Typically in the environmental-surveillance context they will generate an alarm if they detect areas where populations are multi-exposed to environmental pollutants. The new methods appear to be more relevant for multivariate functional data than a multivariate spatial scan statistic approach since the latter would face huge losses of information by summarizing each variable of the data by its average over the time period. Furthermore they also appear to be more relevant than using a univariate functional spatial scan statistic for each variable since this does not allow to take into account the correlations between the variables. \\

\noindent Although they only studied their approach in the univariate functional framework, \cite{wilco_cucala} suggest that the NPFSS can be extended to the multivariate case.
Thus the MDFFSS, the PMFSS and the MRBFSS were compared with the NPFSS in a simulation study. The simulation study highlighted that the performances of all methods decreased with increasing correlation between the variables.
The MRBFSS and the MDFFSS presented higher powers than the PMFSS and the NPFSS whatever the distribution and the correlation between the variables. The PMFSS and the MDFFSS showed the lowest false positive rates. However the MRBFSS presented the highest true positive rates and the PMFSS showed the lowest ones which results in very high F-measures for the MRBFSS and the MDFFSS, which improves the confidence in the clusters detected by these approaches compared to the ones detected by the NPFSS and the PMFSS. Moreover the ones detected by the NPFSS tended to contain more false positives which is less relevant in practice. Indeed in the case of the application of scan statistics to environmental surveillance, having fewer false positives rates is an advantage since the detection of spatial clusters is the starting point for future investigation by environmental experts within the cluster. \\
When the data were far from being normally distributed the performances of the PMFSS decreased as well as the ones of the MDFFSS. However they still maintain very low false positive rates. \\

\noindent
For the sake of brevity we have chosen to apply only the MRBFSS to detect clusters of abnormal values of pollutants concentrations in the \textit{cantons} of \textit{Nord-Pas-de-Calais}, based on the results of the simulation since it shows stable performances whatever the distribution of the variables and the correlation. The method detected a significant most likely cluster in the area of Lille which presents high values of $\text{NO}_2$, $\text{PM}_{10}$ and $\text{PM}_{2.5}$ concentrations. \\

\noindent Remark that we only focused on circular clusters. In the application the maps of pollutants present elongated shapes of high average concentrations especially for $\text{PM}_{10}$ on the coastline, which suggests that other forms of clusters may be relevant to consider in the analysis. As an example \cite{tango} proposed to consider irregularly shaped clusters by considering all sets of sites connected to each other, with a predetermined maximum size. However it should be noted that using this approach generates many more potential clusters than the approach proposed by \cite{spatialscanstat} which drastically increases the computing time. The same disadvantage can be found with the elliptic clusters approach of \citep{elliptic}. However these problems can be overcome with the graph-based clusters proposed by \citep{cucala_graph}. Another possible approach was proposed by \cite{lin} who suggests to regroup the estimated circular clusters to form clusters with arbitrary shapes. \\

\noindent It should also be noted that the application on real data only considered the MLC. It may also be interesting to detect secondary clusters, which can be done by following the approach of \cite{spatialscanstat} who considers also clusters that had a high value of the concentration index ($\mathrm{LH}^{(w)}$ for the PMFSS, $T_{n, \text{max}}^{(w)}$ for the MDFFSS, $W^{(w)}$ for the MRBFSS, and $U(w)$ for NPFSS \citep[see][for details]{wilco_cucala}) and did not cover the MLC. \\

\noindent Finally in the context of spatial epidemiology,  one could imagine case count data collected monthly on spatial units over a long period of time. 
In this context, to detect spatial clusters of diseases with the already existing methods, we often use cumulative incidences, which implies to get only one data per spatial unit.
This induces a huge loss of information, particularly when the incidence curves show high temporal variability.
However we should underline that the NPFSS, the PMFSS, the MDFFSS and the MRBFSS could be applied to count data including the possibility to adjust the analysis for the underlying population.

\newpage
\bibliographystyle{model5-names}
\bibliography{bibliographie}

\appendix

\section{Supplementary materials}
\begin{figure}[h!]
    \centering
    \includegraphics[width = 0.5\textwidth]{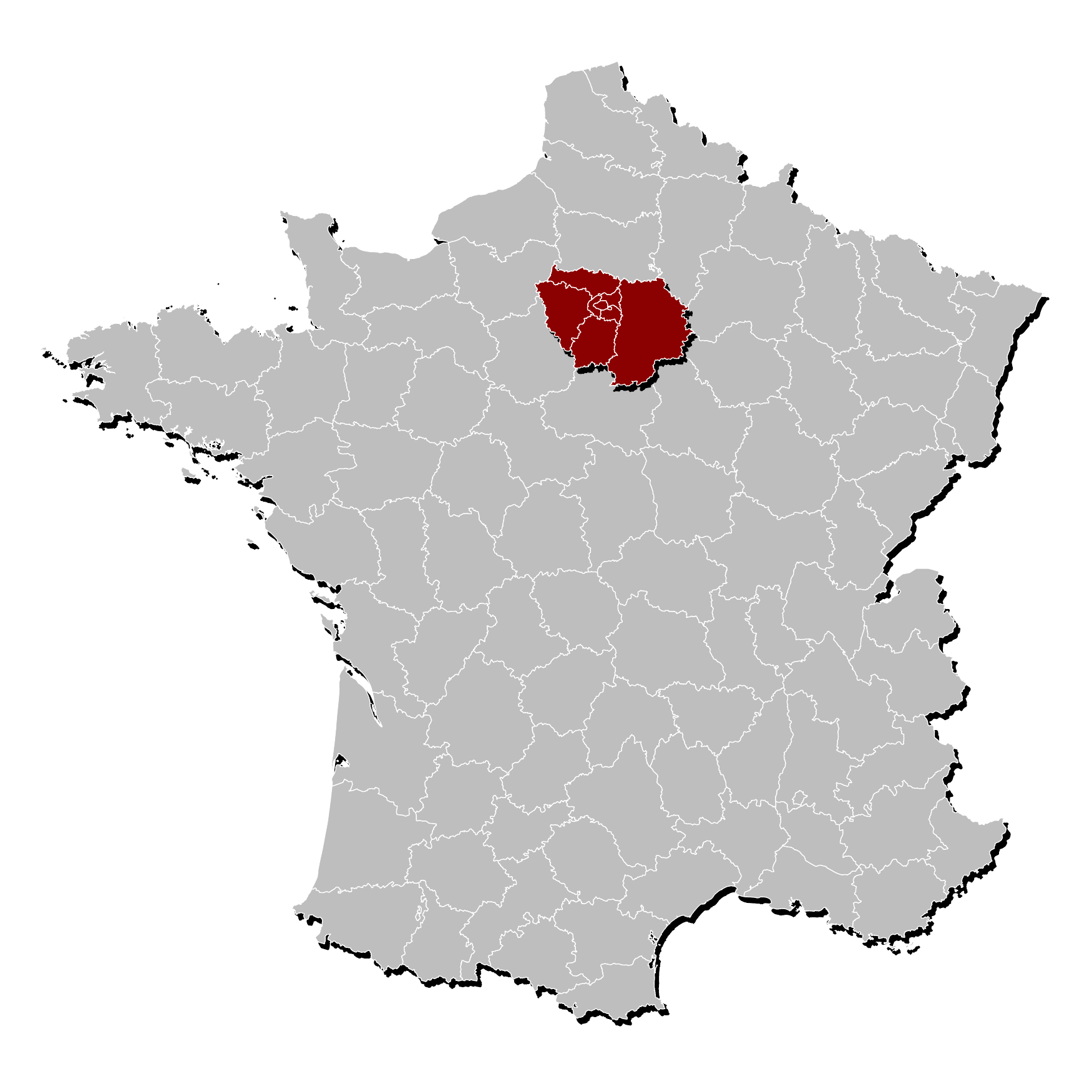}
    \caption{The 94 French \textit{départements} and the spatial cluster (in red) simulated for each artificial dataset.}
    \label{fig:sitesgrid}
\end{figure}

\begin{figure}[h!]
\begin{center}
\includegraphics[width = \linewidth]{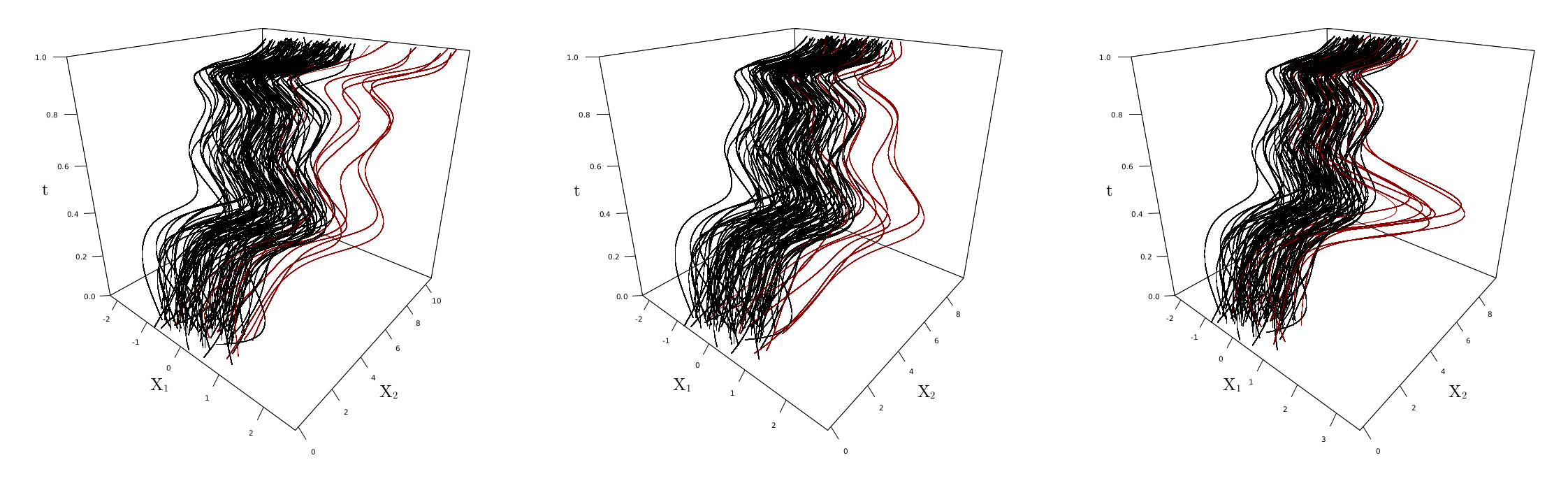}
 \caption{The simulation study: an example of the two components of the data generated for the Gaussian process and $\rho = 0.2$, with $\Delta(t) = \Delta_1(t) = 1.5(t ; t)^\top$ (left panel), $\Delta(t) = \Delta_2(t) = 4(t(1-t) ; t(1-t))^\top$ (middle panel) and $\Delta(t) = \Delta_3(t) = 5( \exp{[ - 100 (t-0.5)^2 ]}/3 ;  \exp{[ - 100 (t-0.5)^2 ]}/3)^\top$ (right panel).
 The red curves correspond to the observations in the cluster.}
  \label{fig:examplesimu}
  
\end{center}
\end{figure}

\noindent Figure \ref{fig:examplesimu} shows an example of the data generated when the $Z_{i,k}$ are Gaussian and $\rho = 0.2$.

\begin{figure}[h!]
\centering
\includegraphics[width = \textwidth]{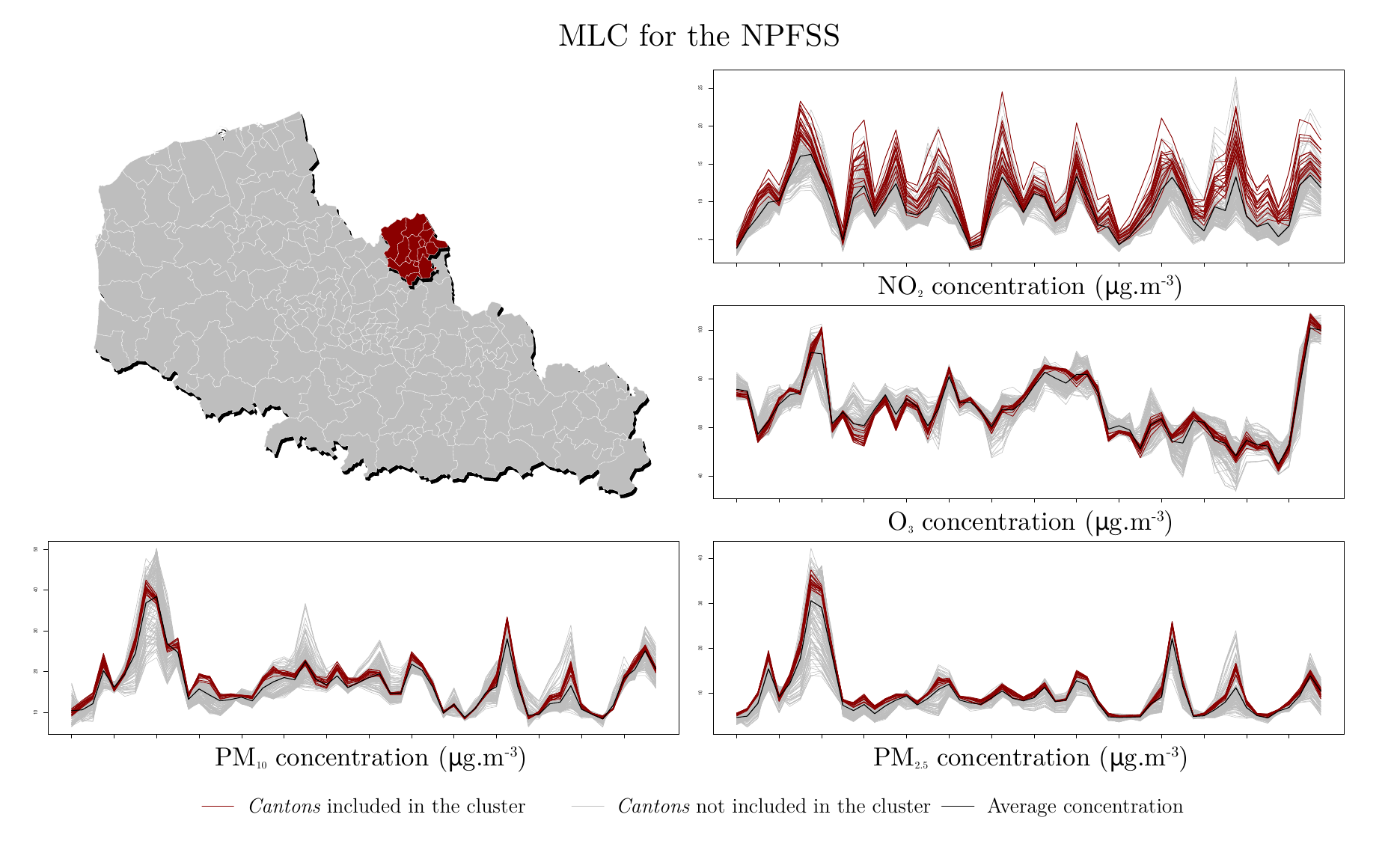}
\end{figure}
\begin{figure}[h!]
\centering
\includegraphics[width = \textwidth]{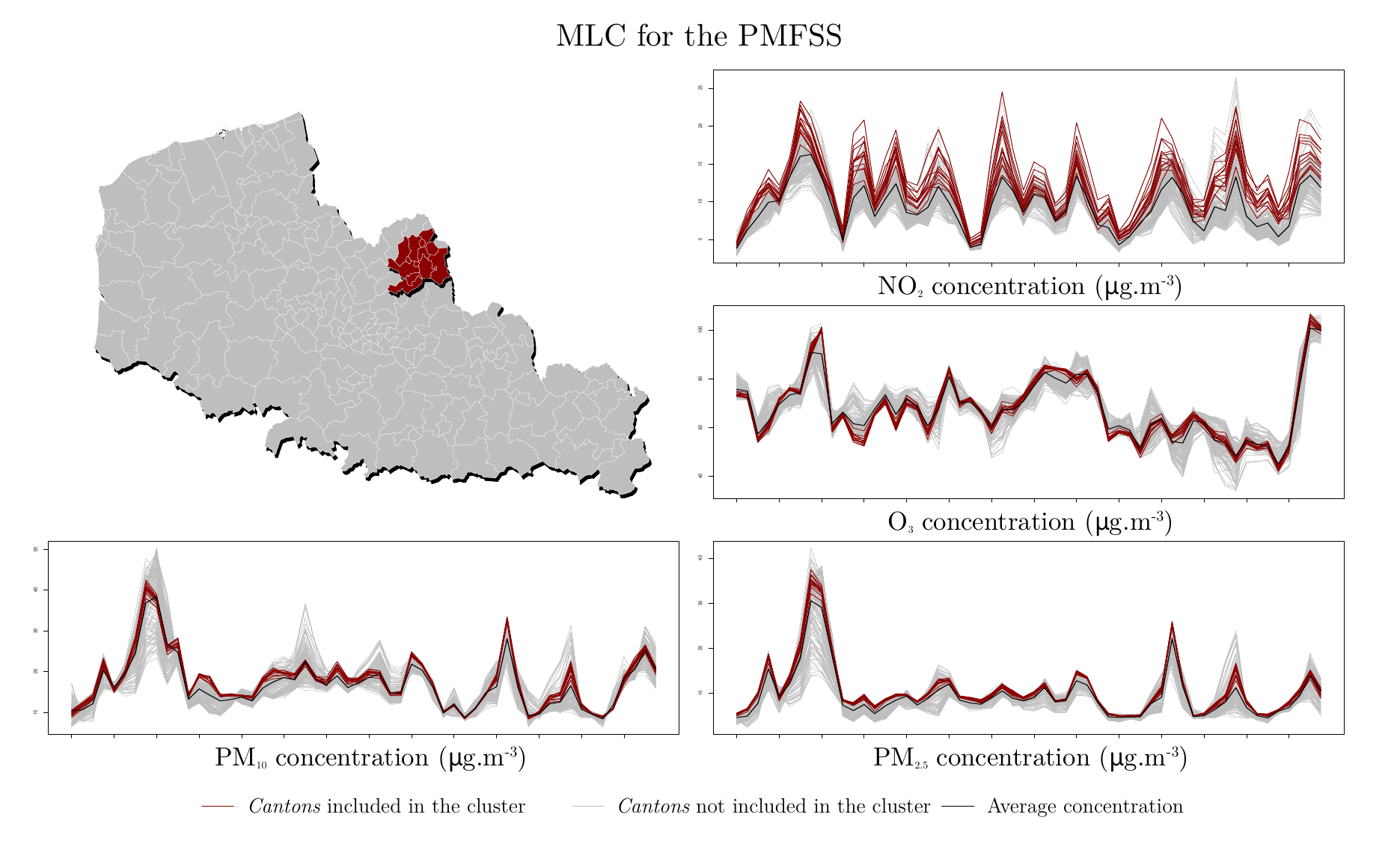}
\end{figure}
\begin{figure}[h!]
\centering
\includegraphics[width = \textwidth]{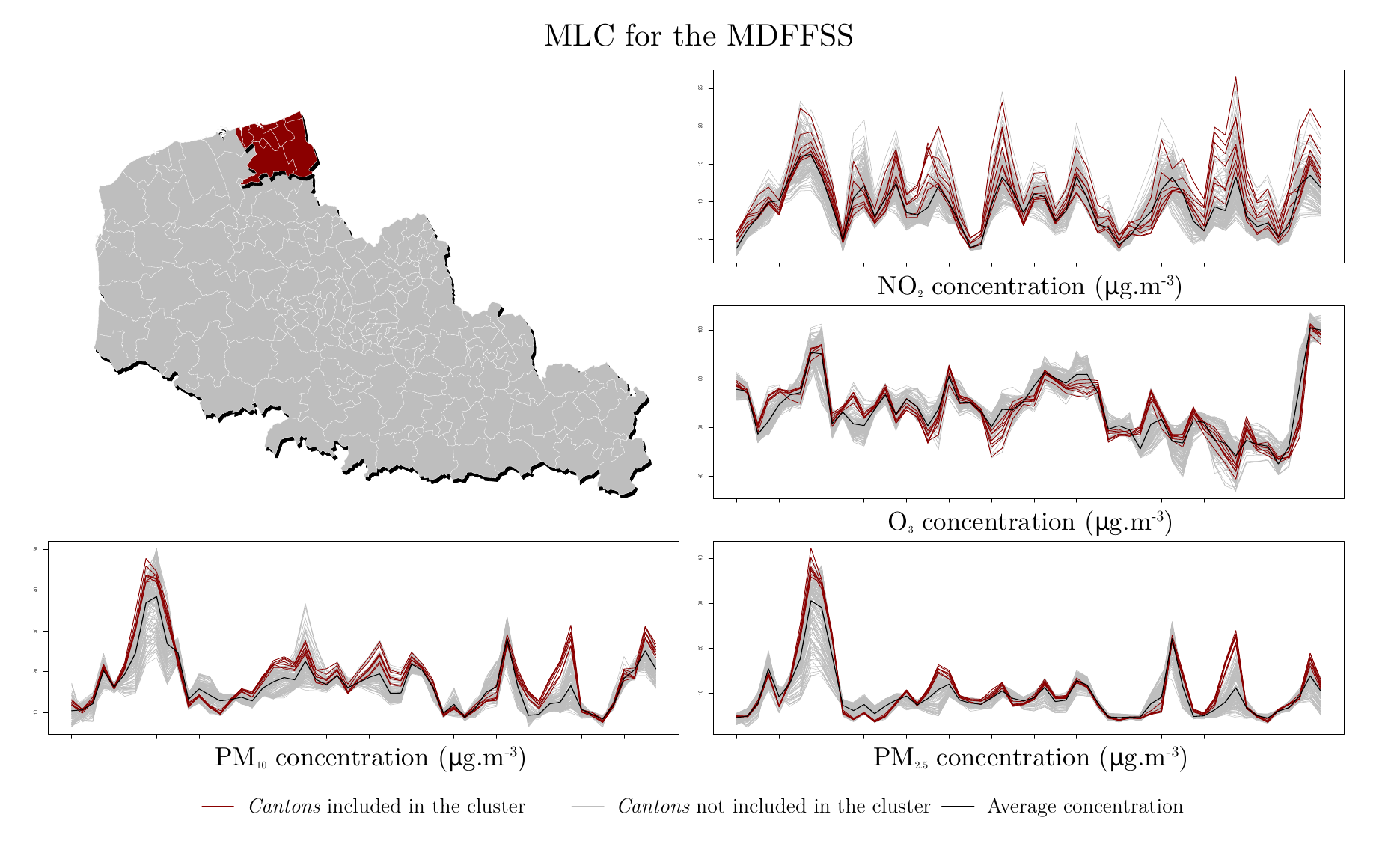}
\caption{Most likely clusters of pollutants ($\text{NO}_2$, $\text{O}_3$, $\text{PM}_{10}$ and $\text{PM}_{2.5}$) concentrations detected by the NPFSS, the PMFSS and the MDFFSS. The daily concentration curves of the pollutants (from May 1, 2020 to June 25, 2020) in each \textit{canton} are presented with colored lines. The black curves are the daily average concentration curves in the \textit{Nord-Pas-de-Calais} (a region in northern France).}
\label{fig:othermethods}
\end{figure}

\begin{table}[htbp]
\centering
\caption{Description of the most likely cluster of pollutants concentrations detected for the NPFSS, the PMFSS and the MDFFSS 
. }

\begin{tabular}{lccc}
\hline
& \multirow{2}{*}{\# \textit{cantons}} & \multirow{2}{*}{Surface} & \multirow{2}{*}{p-value} \\
& & & \\ \hline
\multirow{2}{*}{\textbf{$\Lambda_{\text{NPFSS}}$}} & \multirow{2}{*}{15} & \multirow{2}{*}{308 km$^2$} & \multirow{2}{*}{0.001} \\ 

\multirow{2}{*}{\textbf{$\Lambda_{\text{PMFSS}}$}} & \multirow{2}{*}{13} & \multirow{2}{*}{264 km$^2$} & \multirow{2}{*}{0.001} \\ 

\multirow{2}{*}{\textbf{$\Lambda_{\text{MDFFSS}}$}} & \multirow{2}{*}{7} & \multirow{2}{*}{284 km$^2$} & \multirow{2}{*}{0.001} \\

& & & \\ \hline
\end{tabular}

\label{table:noconstraint}
\end{table}

\noindent Figure \ref{fig:othermethods} presents the most likely clusters obtained with the three other methods. The NPFSS detects exactly the same cluster as the MRBFSS and the most likely cluster for the PMFSS is quite similar to it. The result obtained with the MDFFSS seems at first sight quite surprising. However we only focussed here on the most likely clusters and it was found that this cluster is significant ($\hat{p} = 0.001$) for all methods and that the secondary cluster for the MDFFSS ($\hat{p} = 0.001$) is exactly the MLC of the NPFSS and the MRBFSS. Some characteristics of the detected MLCs are presented in Table \ref{table:noconstraint}.

\end{document}